\documentclass[12pt, prd, showpacs]{revtex4}
\usepackage{amssymb}
\usepackage{amsmath}

\input{tcilatex}

\begin{document}

\title{Pure electric Penrose and super-Penrose processes in the flat
space-time}
\author{O. B. Zaslavskii}
\affiliation{Department of Physics and Technology, Kharkov V.N. Karazin National
University, 4 Svoboda Square, Kharkov 61022, Ukraine}
\affiliation{Institute of Mathematics and Mechanics, Kazan Federal University, 18
Kremlyovskaya St., Kazan 420008, Russia}
\email{zaslav@ukr.net }

\begin{abstract}
Let a particle move in the flat space-time under the action of the electric
potential. If it decays to two fragments, the energy of debris can be larger
than the original energy and, moreover, the efficiency of such a process can
be unbounded. This effect can be considered as a limit of the
Denardo-Ruffini process near the Reissner-Nordstr\"{o}m black hole when the
gravitational constant and/or black hole mass tend to zero. There are
scenarios in which the energy of debris at infinity is much larger than the
initial one. Comparison with other close but distinct effects is discussed.
\end{abstract}

\maketitle

\section{Introduction}

One of impressive effects of gravitation consists in the possibility of the
energy gain due to the process of a particle decay. This can happen in black
hole background, provided the ergoregion (region where negative energies are
possible) exists. When, if particle 0 decays into two fragments 1 and 2 such
that particle 1 acquires energy $E_{2}<0$ and particle 2 returns to
infinity, it becomes possible to have $E_{1}>E_{0}$. This is the so-called
Penrose process (PP) \cite{pen}. The similar effect was noticed for
amplification of waves from rotating bodies \cite{zel1}, \cite{zel2} and
from black holes (superradiance) \cite{st1}, \cite{st2}. Meanwhile, although
both effects are related very \ closely, they are distinct since
superradiance requires the present of the horizon whereas the PP relies on
the existence of the ergoregion only and is possible even without the
horizon \cite{car}.

In recent years, interest to phenomena of such a kind revived. In
particular, it is connected with new findings in high energy collisions near
black holes giving rise to extraction energy from them (see the monograph 
\cite{car} and a recent review \cite{jer}).

The original PP was found for rotating black holes. Meanwhile, later it was
shown that a similar process exists also for static charged black holes, say
the Reissner-Nordstr\"{o}m one \cite{ruf}. In this second case, a black hole
is not rotating. But, anyway, the very presence of a black hole was
essential for the effect to occur. In the present work, we draw attention to
the fact that the PP in the version of Ref. \cite{ruf} persists even in the
limit when the gravitational constant $G\rightarrow 0$ (equivalently, the
black hole mass $M\rightarrow 0$ and the metric becomes flat)!
Alternatively, one can start from the flat space-time from the very
beginning.

Another aim of the present work is to evaluate the efficiency $\eta $ of the
PP. Near rotating uncharged black holes it is quite modest \cite{jer}, \cite%
{pir3}, \cite{p} - \cite{z}. However, it was noticed earlier that near
charged black holes electromagnetic interaction favours PP \cite{d1}, \cite%
{d2}. Moreover, it was shown in \cite{rn} that in the collisional version of
the PP in the test particle approximation $\eta $ \ can grow without bound.
By definition, this is the so-called super-Penrose process (SPP). This
result was obtained in the situation when the energy $E_{c.m.}$ in the
centre of mass frame of two particles also grows without bound \cite{jl}
that represents a counterpart of the Ba\~{n}ados-Silk-West (BSW) effect
found for rotating black holes \cite{ban}. In the present work, I show that
SPP is also possible in the situation when $E_{c.m.}$ is finite and a black
hole is absent at all.

It is worth stressing that, in spite of similarities between the rotating
and charged versions of the effects under discussion, there is also big
difference between them not only in the conditions under which they occur
(see above) but also in the results. For the rotating uncharged black hole
case the SPP is impossible (see \cite{fraq} \ and references therein). It is
possible near another strongly gravitating objects such as naked
singularities \cite{inf}, \cite{ourwald} or wormholes \cite{worm}. But we
will see that for the charged case the SPP is not only possible but this
happens even in the flat space-time.

Thus I make accent on two properties of the effect under discussion that
were absent in the rotating case: (i) the existence of the flat space-time
limit of PP in the black hole background, (ii) the existence of the SPP with
a finite $E_{c.m.}$

After the preprint version of the present work had been finished, I became
aware that the possibility of the pure electric PP in the flat space-time
was considered (probably, for the first time) in the work \cite{dad} where
several numeric examples were demonstrated. Our consideration is more
general and all the results are obtained in the analytical form. Also, we
make the accent on the relevance of the SPP that was not discussed in \cite%
{dad}.

\section{Basic equations}

Let us consider radial motion of a particle with a mass $m$ and electric
charge $q$ in the flat space-time in the electric potential $\varphi $.
Then, the equations of motion read%
\begin{equation}
m\dot{t}=X\text{,}  \label{tx}
\end{equation}%
\begin{equation}
m\dot{r}=\sigma P\text{, }
\end{equation}%
where $\sigma =\pm 1$ depending on the direction of motion,%
\begin{equation}
X=E-q\varphi \text{, }  \label{X}
\end{equation}%
\begin{equation}
P=\sqrt{X^{2}-m^{2}-\frac{L^{2}}{r^{2}}}\text{,}  \label{z}
\end{equation}%
dot denotes derivative with respect to the proper time $\tau $, $L$ being
the angular momentum. The so-called forward-in-time condition $\dot{t}>0$
requires%
\begin{equation}
X>0\text{.}  \label{ft}
\end{equation}

We assume that $\varphi \geq 0$. In the static situation, the natural gauge
freedom reduces to the choice of the constant in $\varphi $. However, the
energy can be shifted to the same constant, so $X$ remains gauge-invariant.
To fix the gauge, we imply that at infinity $\varphi \rightarrow 0$, $%
E\rightarrow \frac{m}{\sqrt{1-V^{2}}}$, where $V$ is the velocity \ (the
speed of light $c=1$). We assume the Coulomb potential $\varphi =\frac{Q}{r}$
with $Q>0$.

Let particle 0 move from large $r$ towards the centre and decays in some
point $r=r_{0}$ to particles 1 and 2. Then, it follows from the conservation
laws that%
\begin{equation}
E_{0}=E_{1}+E_{2}\text{,}  \label{e}
\end{equation}

\begin{equation}
q_{0}=q_{1}+q_{2}\text{,}  \label{q}
\end{equation}%
\begin{equation}
L_{0}=L_{1}+L_{2},
\end{equation}%
\begin{equation}
-P_{0}=\sigma _{2}P_{2}+\sigma _{1}P,  \label{pr}
\end{equation}
It follows from (\ref{e}) and (\ref{q}) that%
\begin{equation}
X_{0}=X_{1}+X_{2}\text{.}  \label{x}
\end{equation}

\section{Kinematics of motion}

It is convenient to introduce notation%
\begin{equation}
U(r)=\sqrt{m^{2}+\frac{L^{2}}{r^{2}}}+q\frac{Q}{r}\text{.}
\end{equation}%
Then, taking into account that $X>0$, we see that the allowed region of
motion corresponds to%
\begin{equation}
E\geq U(r)\text{.}  \label{eu}
\end{equation}

The properties of particle motion depends on the sign of the electric charge.

\subsection{$q\geq 0$}

Then, $U$ is monotonically decreasing from $U(0)=\infty $ to $U(\infty )=m$.
There is only 1 turning point $r_{0}$ where $E=U(r_{0})$. Any particle
moving from infinity bounces back in this point and returns to infinity.

\subsection{$q=-\left\vert q\right\vert <0$}

If $\alpha \equiv \left\vert q\right\vert Q>\left\vert L\right\vert $, $U(r)$
is monotonically increasing function, $U(0)=-\infty $, $U(\infty )=m$. If $%
E\geq m$, there are no turning points, a particle escapes or falls in the
centre depending on the direction of motion. If $E<m$, there is one turning
point. Particle cannot escape to infinity and falls in the centre $r=0$.

If 
\begin{equation}
\alpha \leq \left\vert L\right\vert ,  \label{L}
\end{equation}
there is one minimum of the potential $U(r)$ in the point 
\begin{equation}
r_{m}=\frac{\left\vert L\right\vert }{m\alpha }\sqrt{L^{2}-\alpha ^{2}}\text{%
.}
\end{equation}%
Then,%
\begin{equation}
U(r_{m})=\frac{m}{\left\vert L\right\vert }\sqrt{L^{2}-\alpha ^{2}}\geq 0%
\text{,}
\end{equation}%
$U(0)=\infty $, $U(\infty )=m$.

Thus if $0<E<m$ and (\ref{L}) holds, a particle oscillates inside a
potential well.

\section{Negative energy states}

The positivity of $X$ is compatible with $E<0$, provided $q<0$. The boundary
of the corresponding zone $r\leq r_{erg}$ (generalized ergoregion) is given
by $U(r_{erg})=0$, whence%
\begin{equation}
r_{erg}=\frac{\sqrt{\alpha ^{2}-L^{2}}}{m}\text{.}  \label{rerg}
\end{equation}%
We see that $r_{erg}$ monotonically decreases with $\left\vert L\right\vert $%
. If $L=0$,%
\begin{equation}
r_{erg}=\frac{\left\vert q\right\vert Q}{m}\text{.}  \label{rergo}
\end{equation}

It follows from (\ref{X}), (\ref{ft}) that such states are forbidden if $q>0$%
. It is seen from (\ref{rerg}) that they are also impossible if $q<0$ but $%
\left\vert q\right\vert Q<\left\vert L\right\vert $. Thus, the negative
energy states are allowed only for 
\begin{equation}
q<-\frac{\left\vert L\right\vert }{Q}
\end{equation}%
or, equivalently, 
\begin{equation}
\alpha >\left\vert L\right\vert  \label{a}
\end{equation}%
If $L=0$ (radial motion) negative energies are possible for any $q<0$.

As the conditions (\ref{L}), (\ref{a}) are mutually inconsistent, there are
no oscillating states with $E<0$: such a particle falls in the centre. This
is an exact counterpart of what happens to the negative energy states in the
gravitational case \cite{ghe}, \cite{one}.

Below, we consider different scenarios of splitting of the original particle.

\section{ Scenario 1: pure radial motion}

We assume that particle 2 continues to move towards the center, so $\sigma
_{2}=-1.$ If $\sigma _{1}=-1$, particle 1 continues to move to the centre as
well and only afterwards bounces back from the turning point. If $\sigma
_{1}=+1$, particle 1 moves to infinity immediately after collision. For
simplicity, we take $L_{0}=L_{1}=L_{2}=0$. It is supposed that all masses $%
m_{i}$ and charged $q_{i}$ (i=0,1,2) are fixed.

Then, we have one equation for unknown $X_{1}$:%
\begin{equation}
P_{0}+\sigma _{1}\sqrt{X_{1}^{2}-m_{1}^{2}}=\sqrt{(X_{0}-X_{1})^{2}-m_{2}^{2}%
}\text{.}  \label{xx}
\end{equation}

It follows from it and from (\ref{x}) that%
\begin{equation}
X_{1}=\frac{1}{2m_{0}^{2}}(X_{0}\Delta _{+}+\delta P_{0}\sqrt{D})\text{,}
\label{x1d}
\end{equation}%
\begin{equation}
X_{2}=\frac{1}{2m_{0}^{2}}(X_{0}\Delta _{-}-\delta P_{0}\sqrt{D}),
\label{x2d}
\end{equation}%
where $\delta =\pm 1$, all quantities are taken in the point of decay $%
r=r_{0}$,%
\begin{equation}
\Delta _{\pm }=m_{0}^{2}\pm (m_{1}^{2}-m_{2}^{2}),
\end{equation}%
\begin{equation}
D=\Delta _{+}^{2}-4m_{0}^{2}m_{1}^{2}=\Delta _{-}^{2}-4m_{0}^{2}m_{2}^{2}%
\text{.}
\end{equation}

It is easy to check that the first term in eqs. (\ref{x1d}) and (\ref{x2d})
is always larger than the second one. Therefore, eq. (\ref{ft}) for each
particle entails $\Delta _{+}>0$, $\Delta _{-}>0$, whence%
\begin{equation}
m_{0}^{2}>\left\vert m_{1}^{2}-m_{2}^{2}\right\vert \text{.}  \label{m012}
\end{equation}%
Also, the condition $D\geq 0$ gives us $\Delta _{+}\geq 2m_{0}m_{1}$, $%
\Delta _{-}\geq 2m_{0}m_{2}$, whence%
\begin{equation}
(m_{0}-m_{1})^{2}\geq m_{2}^{2}
\end{equation}%
and%
\begin{equation}
(m_{0}-m_{2})^{2}\geq m_{1}^{2}\text{.}
\end{equation}

These inequalities are consistent with each other if%
\begin{equation}
m_{0}\geq m_{1}+m_{2}\text{.}  \label{m+}
\end{equation}

In particular, it follows from (\ref{m+}) that the case $m_{0}=0$ is
impossible, so, for instance, a photon cannot turn into the
electron-positron pair. This fact is well known and immediately; follows
from the consideration of the process in the centre of mass of such a pair.

The inequality (\ref{m+}) has a simple meaning. In the centre of mass frame
(the frame comoving with particle 0) the conservation of energy gives us%
\begin{equation}
m_{0}=\gamma _{1}m_{1}+\gamma _{2}m_{2}\text{,}
\end{equation}%
where $\gamma _{1,2}$ are corresponding Lorentz factors of particle 1 and 2.
As $\gamma _{1,2}\geq 1$, condition (\ref{m+}) should be fulfilled.

By substitution of (\ref{x1d}), (\ref{x2d}) back into (\ref{xx}) one can
check that the following cases can be realized: $(+,+,+,-)$,$~(+$, $-$, $-$,
\thinspace $+)$, $(-,$ $+,-,$ $+)$ and $(-,-,+,-)$. Here we used notations $%
(\sigma _{1},\varepsilon _{1},\varepsilon _{2}$,$\delta )$, where $%
\varepsilon _{1}=+1$ if $\Delta _{+}^{2}\geq 4m_{1}^{2}X_{0}^{2}$ and $%
\varepsilon _{1}=-1$ if $\Delta _{+}^{2}<4m_{1}^{2}X_{0}^{2}$, $\varepsilon
_{2}=+1$ if $\Delta _{-}^{2}\geq 4m_{2}^{2}X_{0}^{2}$ and $\varepsilon
_{2}=-1$ if $\Delta _{-}^{2}<4m_{2}^{2}X_{0}^{2}$.

Below, we consider different particular scenarios.

\subsection{Scenario 1a: $X_{0}(r_{0})\gg m_{0,1,2}$, $E_{0}$}

According to (\ref{X}), the quantity $X_{0}$ can be very large either due to
large $E_{0}$ or large and negative $q_{0}$. The first possibility is not
very interesting since one must take very large energy from the very
beginning, then it is not surprising if an outcome will be also large.
Therefore, we assume that $E_{0}=O(m_{0})$ is finite but $q_{0}\gg m_{0}$.

Here, only subcase $(+,-,-,+)$ can be realized. In doing so, $P_{0}\approx
X_{0}$ is very large as well, 
\begin{equation}
X_{1}\approx \frac{X_{0}(\Delta _{+}+\sqrt{D})}{2m_{0}^{2}}\text{,}
\label{x1l}
\end{equation}%
\begin{equation}
X_{2}\approx \frac{X_{0}(\Delta _{-}-\sqrt{D})}{2m_{0}^{2}}.  \label{xl2}
\end{equation}

We see that $X_{1}$ is also very large. To avoid a problem with the
evaluation of electric interaction between particles 1 and 2 (that is not
related directly to the PP or SPP and only obscures the whole picture) one
can put $q_{1}=0$, $q_{2}=q_{0}=-\left\vert q_{0}\right\vert .$

Let all masses have the same order. Then, it follows from (\ref{X}) and (\ref%
{x1l}) that%
\begin{equation}
E_{1}=X_{1}=O(X_{0})\sim \left\vert q_{0}\right\vert \varphi (r_{0})\gg
E_{0}.
\end{equation}

Thus%
\begin{equation}
\eta =\frac{E_{1}}{E_{0}}\gg 1,
\end{equation}%
so we deal with the SPP ! In doing so, $E_{2}$ should be large negative$,$
we have from (\ref{xl2})%
\begin{equation}
E_{2}=X_{2}-\left\vert q_{0}\right\vert \varphi (r_{0})\approx -\frac{%
\left\vert q_{0}\right\vert \varphi (r_{0})}{2m_{0}^{2}}%
(m_{0}^{2}+m_{1}^{2}-m_{2}^{2}+\sqrt{D})\text{.}
\end{equation}

According to (\ref{m012}), the expression in parentheses is positive, so
indeed $E_{2}<0$.

To simplify formulas, we can consider the case 
\begin{equation}
m_{1}=m_{2}\equiv m,m_{0}=2m.  \label{2m2}
\end{equation}%
Then, $D=0$ and one finds from (\ref{x1d}), (\ref{x2d}):%
\begin{equation}
X_{1}=X_{2}=\frac{X_{0}}{2}\text{, }E_{1}=\frac{E_{0}+\left\vert
q_{0}\right\vert \varphi (r_{0})}{2}\text{,}  \label{x1}
\end{equation}%
\begin{equation}
E_{2}=\frac{E_{0}-\left\vert q_{0}\right\vert \varphi (r_{0})}{2},
\label{e2}
\end{equation}%
\begin{equation}
\eta =\frac{1}{2}+\frac{\left\vert q_{0}\right\vert \varphi (r_{0})}{2E_{0}}%
\text{.}  \label{nu}
\end{equation}

If $\left\vert q_{0}\right\vert \varphi (r_{0})>E_{0}$, the PP occurs. If $%
\left\vert q_{0}\right\vert \varphi (r_{0})\gg E_{0}$ we obtain the SPP.

As an initial particle was overcharged and it is neutral particle that
returns to infinity, one can say that the initial electric charge converts
into energy.

\subsection{Scenario 1b: Finite $X_{1}$}

There is also another option. Let $q_{0}$, $X_{0}$ have the same order as
all masses and $E_{0}$. Then, $X_{1}$ is also finite and has the same order
according to (\ref{x1d}).\ If $\left\vert q_{1}\right\vert \varphi (r_{0})$ $%
\gg E_{0}$ we again obtain that $\eta \sim \frac{\left\vert q_{1}\right\vert
\varphi (r_{0})}{E_{0}}\gg 1$.

It is worth noting that in any case, the energy $E_{c.m.}$ in the centre of
mass frame of particles 1 and 2 is finite, $E_{c.m.}=m_{0}$.

\section{ Scenario 2: splitting in the turning point}

Let us we consider the case with nonzero angular momenta. We assume that
splitting of an original particle 0 occurs in the turning point $r_{0}$
where $P_{0}=0$, so%
\begin{equation}
X_{0}=E_{0}-q_{0}\varphi (r_{0})=\sqrt{m_{0}^{2}+\frac{L_{0}^{2}}{r_{0}^{2}}}%
\text{.}  \label{x0}
\end{equation}%
Then, it is easy to obtain from (\ref{pr}) that%
\begin{equation}
m_{0}^{2}+m_{1}^{2}-m_{2}^{2}+\frac{2L_{0}L_{1}}{r_{0}^{2}}=2X_{1}\sqrt{%
m_{0}^{2}+\frac{L_{0}^{2}}{r_{0}^{2}}}
\end{equation}

If%
\begin{equation}
m_{0}=m_{1}+m_{2}  \label{m0}
\end{equation}%
we return to eq. (12) of \cite{dad} in somewhat different notations
(however, indices 1 and in eq. 12 of \cite{dad} should be interchanged):%
\begin{equation}
X_{1}=\frac{1}{X_{0}}(m_{0}m_{1}+\frac{L_{0}L_{1}}{r_{0}^{2}}).  \label{12}
\end{equation}%
In a similar way,%
\begin{equation}
X_{2}=\frac{1}{X_{0}}(m_{0}m_{2}+\frac{L_{0}L_{2}}{r_{0}^{2}})\text{.}
\end{equation}

Let us consider case (\ref{2m2}) with $L_{1}=L_{2}=\frac{L_{0}}{2}$. Then,%
\begin{equation}
X_{1}=X_{2}=\frac{X_{0}}{2}\text{.}
\end{equation}%
Eqs. (\ref{2m2}) - (\ref{nu}) hold true.\ However, in the aforementioned
equations it was implied that all angular momenta vanish. Now, this is not
so. It follows from (\ref{x0}) that for large negative $q_{0}$ we have%
\begin{equation}
\left\vert q_{0}\right\vert \varphi (r_{0})\approx \frac{\left\vert
L_{0}\right\vert }{r_{0}}\gg m_{0}\text{.}
\end{equation}%
We can take $q_{1}=0$, $q_{0}<0$, $E_{0}\sim m_{0}$. Then, $X_{1}\sim
\left\vert q_{0}\right\vert \varphi (r_{0})\gg E_{0}$, so the efficiency $%
\eta \gg 1$ and we are faced with the SPP\ again.

\section{Energy at infinity and energy in the centre of mass frame}

There is one more interesting point in this context. In \cite{rn}, \cite{esc}
it was assumed that the energy in the $E_{c.m.}$ centre of mass frame of two
particles is unbounded that produces the analogue of the Ba\~{n}%
ados-Silk-West (BSW) effect \cite{ban} near nonrotating charged black holes 
\cite{jl}. It was shown for rotating space-times \cite{inf}, \cite{ourwald}
that the energy $E_{1}$ of debris at infinity can be unbounded only if $%
E_{c.m.}$ is unbounded as well, otherwise $E_{1}$ is bounded. In particular,
this follows from the Wald inequalities \cite{wald}, as found in \cite%
{ourwald}.

Meanwhile, now the situation is different. Repeating step by step derivation
of the Wald inequalities \cite{wald}, it is easy to understand that the only
difference is that one should replace the energies $E_{i}$ with $X_{i}$
(generalized momenta). Then, in the flat space-time the counterpart of eq.
(1) from \cite{wald} reads%
\begin{equation}
\gamma _{1}[\frac{X_{0}}{m_{0}}-v_{1}\sqrt{\left( \frac{X_{0}}{m_{0}}\right)
^{2}-1}]\leq \frac{X_{1}}{m_{1}}\leq \gamma _{1}[\frac{X_{0}}{m_{0}}+v_{1}%
\sqrt{\left( \frac{X_{0}}{m_{0}}\right) ^{2}-1}]\text{,}
\end{equation}%
where $\gamma =\frac{1}{\sqrt{1-v_{1}^{2}}}$, $v_{1}$ is the value of the
three-velocity that particle 1 has in the frame comoving to particle 0. Let,
for simplicity, $m_{1}=m_{2}=m$. Then, the conservation of energy on the
centre of mass frame gives us $m_{0}=2\gamma _{1}m$, whence%
\begin{equation}
X_{0}-v_{1}\sqrt{X_{0}^{2}-m_{0}^{2}}\leq 2X_{1}\leq X_{0}+v_{1}\sqrt{%
X_{0}^{2}-m_{0}^{2}}\text{.}
\end{equation}

If $X_{0}\gg m_{0}$, $X_{1}=O(X_{0})$. This corresponds to scenario 1a and
scenario 2. If $X_{0}$ is bounded, the quantity $X_{1}$ is bounded as well.
Then, for a given $X_{1}$, the energy can become unbounded due to unbounded $%
q_{1}$. This corresponds to scenario 1b. In all cases, $E_{c.m.}=m_{0}$
remains bounded.

\section{PP, SPP\ and other effects}

In general, there are several versions of related but different effects that
show some similarities but do not coincide. This group contains (i) the
Schwinger effect of pair creation \cite{pair}, (ii) the "classical" PP near
rotating black holes, (iii) amplification of waves by a cylinder, (iv) the
PP and SPP near charged nonrotating black holes. The process under
discussion to some extent resembles process (i) but there are also important
differences here. The Schwinger effect effects is purely quantum and does
not exist in the classical limit. It is also nonlocal since there is a
minimum separation $l_{\min }$ between the electron and positron created in
the electric field \thinspace $\mathcal{E}$, $l_{\min }=\frac{\left\vert
e\right\vert \mathcal{E}}{2m}$, $\left\vert e\right\vert $ is the value of
the electron charge. In the Schwinger effect, the initial state is vacuum.
But in our example there is an initial particle with $q_{0}$ that, in
general, can be nonzero. From another hand, the results described above
include the case when one of particle is neutral. This is impossible for the
Schwinger effect. The results obtained in the present work admit
straightforward generalization to the collisional version (particle 1 and 2
collide producing particles 3 and 4) whereas it makes no sense for the
Schwinger effect.

In case (ii) the efficiency of the process is essentially restricted. The
collisional version improves the situation but anyway $\eta $ remains
bounded (see \cite{jer}, \cite{fraq} and references therein). Meanwhile, now
it remains unbounded. Process (iii) is possible in the flat space-time \cite%
{zel1}, \cite{zel2} as well as the present one but physical sources are
different. In this sense, relation between (ii) and the present work to some
extent resembles the relation between \cite{pen} and \cite{ruf}. The SPP
near charged nonrotating black holes was already pointed out in \cite{rn}
(confirmed in \cite{esc}). But it was essential in the aforementioned works
that the energy in the centre of mass $E_{c.m.}$ grew unbounded. Meanwhile,
now $E_{c.m.}$ remains finite.

Thus \ we see that in spite of some more or less closed similarities, we
deal with different effects. And, it is interesting by itself that we
obtained the PP and SPP processes in the flat space-time from a black hole
perspective, taking the flat limit of the previously known effect \cite{ruf}.

The obtained results give formally unbounded energy gain for $E_{1}$ but it
requires an infinite charge. Obviously, in real nature the electric charge
is bounded due to some other factors, first of all due to the inequality $%
\left\vert q\right\vert <A\left\vert e\right\vert $ relevant in quantum
electrodynamics ($e$ is the electric charge of the electron, $A\approx 170$ 
\cite{zp}). But even with such a restriction, $E\sim A\left\vert
e\right\vert \varphi (r_{0})$ can be quite large due to large $A$. Thus the
efficiency of the process can be very large in a classical theory, being
restricted by quantum processes, the size of nuclei, etc. \ It is also worth
mentioning that all results are obtained in the test particle approximation,
with backreaction neglected. If the charge $Q>0$, particles with $q_{2}<0$
will attract to the centre and screen the initial charge that will change
the whole picture. We do not go here into such details. In our view, the
very fact that we obtain SPP\ in the region of validity of classical theory
at least in the test particle approximation is interesting by itself.

In general, there is also an additional subtle point. Present consideration
neglects direct electric interaction between particles 1 and 2. Strictly
speaking, instead of considering point-like particles one must introduce
some effective radius of interaction $a$. Happily, this point is irrelevant
for us in main scenarios 1a and 2 since we managed to build examples in
which particle 1 is neutral.

It is clear form the derivation that the process under discussion occurs
also for any regular curved space-time, say in the black hole background.
For simplicity, let a particle move radially, so $L=0$. Then, instead of (%
\ref{z}), we would have $P=\sqrt{X^{2}-m^{2}N^{2}}$. Thus, decay in an
intermediate point with $N\sim 1$ would give qualitatively the same result.
It completely agrees with that of \cite{ruf}. According to eq. (6) of \cite%
{ruf}, the boundary of the region with possible $E\leq 0$ is given for
radial motion by the expression%
\begin{equation}
\frac{Q\left\vert q_{2}\right\vert }{r_{erg}}=m_{2}N\text{,}
\end{equation}%
where $N$ is the lapse function of the Reissner-Nordstr\"{o}m metric. In the
flat space-time limit, $N\rightarrow 1$, and we return to (\ref{rergo}).
Meanwhile, it is worth stressing that not only the PP is possible for the
decay with $r<r_{erg}$ as pointed out in \cite{ruf} but also the SPP.

As far as the role of rotation is concerned, it creates obstacle to the SPP.
Indeed, the radial momentum\ $P=\sqrt{X^{2}-N^{2}(m^{2}+\frac{L^{2}}{r^{2}})}
$, where now $X=E-\omega L$, $\omega =-g_{0\phi }/g_{\phi \phi }>0$, $L$
being an angular momentum \cite{z}. For a given finite $X,$ large positive $%
E $ requires large positive $L$. Then, if particle decay occurs at some
intermediate point where $N\sim 1$, the squared radial momentum $P^{2}$
becomes negative since the negative term $\frac{L^{2}}{r^{2}}$ is very large
whereas the positive $X^{2}$ term is finite. This does not happen when $%
\omega =0$ but the electric charge $q$ is present. Then, $q$ enters $P$
indirectly, via $X$ only and there are no large negative terms inside the
square root. In this respect, it is the electric charge (but not rotation)
which favours the existence of the SPP.

\section{Summary}

We obtained several scenarios in which PP and even SPP\ can occur. What are
the key physical reasons for this? First of all, this is possibility of
having negative energies. It is absent for free particles but becomes
possible in the external electric field. In this case, it is the
gauge-invariant combination $X=E-q\varphi $ that enters the formulas.
According to (\ref{tx}), it measures the rate with which the time changes
along the trajectory and should be positive (the forward-in-time condition).
Meanwhile, positivity of $X$ is compatible with negativity of $E$, provided $%
q\varphi <0$ (say, $\varphi >0$ and $q<0$) and is large enough. Thus there
are two relevant conditions (i) $E<0$, (ii) the validity of the
forward-in-time condition. It is the combination of (i) and (ii) that makes
the PP possible. The existence of region with $E<0$ (generalized ergoregion)
was discussed in black hole physics \cite{ruf} as a counterpart to the
standard PP for rotating black holes \cite{pen}. However, as we see, the
presence of a black hole is not necessary, the effect remains even if
gravity effects vanish.

As far as the SPP is concerned, a serious obstacle here is connected with
the Wald inequalities \cite{wald} that restrict the energy for the rotating
metrics and uncharged particles. To overcome these restriction, one is led
to quite sophisticated collisional scenarios in a strong gravitational field 
\cite{inf}, \cite{ourwald}. But now, because it is $X$ (but not only $E$
itself) enters the formulas, bounded $X$ are quite compatible with very
large $E$, provided $\left\vert q\right\vert $ is also very large.

Thus we showed that some interesting effects that are usually considered as
inherent to objects in a strong gravitation field, can occur even in the
flat space-time. As follows from \cite{dad} and the present work, this is
true for the PP process. And, as found in the present work, this is valid
for SPP as well. We considered the general scheme and obtained the results
that follow from the conservation law, without pretending to concrete
applications. We hope that application of the ideas of \cite{dad} and of the
present work to more concrete systems (say, in nuclear or atomic physics)
would be of interest on its own displaying close analogies between
gravitation and other branches of physics.

\begin{acknowledgments}
I thank Yu. P. Stepanovsky for interesting discussion. This work was funded
by the subsidy allocated to Kazan Federal University for the state
assignment in the sphere of scientific activities. O. Z. also thanks for
support SFFR, Ukraine, Project No. 32367.
\end{acknowledgments}

\end{document}